\documentclass[twocolumn]{aastex62}
\usepackage{graphicx}
\usepackage{dcolumn}
\usepackage{bm}
\usepackage{epstopdf,color,verbatim}
\usepackage{hyperref}
\graphicspath{{./}{figures/}}

\received{January 1, 2018}
\revised{January 7, 2018}
\accepted{\today}
\submitjournal{ApJ}

\shorttitle{Pair-Compton cascades with TPP}
\shortauthors{Petropoulou et al.}

\newcommand{\eqb}{\begin{eqnarray}}
\newcommand{\eqe}{\end{eqnarray}}
\newcommand{\sth}{\sigma_{\rm T}}
\newcommand{\gmax}{\gamma_{eq}}
\newcommand{\xav}{\langle \epsilon \rangle}
\newcommand{\sect}[1]{Section~\ref{sec:#1}}

\newcommand{\eq}[1]{Equation~(\ref{eq:#1})}
\newcommand{\fign}[1]{Figure~\ref{fig:#1}}

\begin{document}

\title{Inverse Compton Cascades in Pair-Producing Gaps: \\ 
Effects of Triplet Pair Production}

\correspondingauthor{Maria Petropoulou}
\email{m.petropoulou@astro.princeton.edu}

\correspondingauthor{Yajie Yuan}
\email{yajiey@astro.princeton.edu}

\author[0000-0001-6640-0179]{Maria Petropoulou}
\affil{Department of Astrophysical Sciences, Princeton University \\
4 Ivy Lane, Princeton, NJ 08544, USA}

\author{Yajie Yuan}
\affil{Department of Astrophysical Sciences, Princeton University \\
4 Ivy Lane, Princeton, NJ 08544, USA} 

\author{Alexander Y. Chen}
\affil{Department of Astrophysical Sciences, Princeton University \\
4 Ivy Lane, Princeton, NJ 08544, USA} 

\author{Apostolos Mastichiadis}
\affil{Department of Physics, National \& Kapodistrian University of Athens \\
Panepistimiopolis, 15784 Zografos
Greece} 


\begin{abstract}
Inverse Compton-pair cascades are initiated when gamma-rays are absorbed on an ambient soft photon field to produce relativistic pairs, which in turn up-scatter the same soft photons to produce more gamma-rays. If the Compton scatterings take place in the deep Klein-Nishina regime, then triplet pair production ($e\gamma_b \rightarrow ee^{+}e^{-}$) becomes relevant and may even regulate the development of the cascade. 
We investigate the properties of pair-Compton cascades with triplet pair production in accelerating gaps, i.e., regions with an unscreened electric field.  Using the method of transport equations for the particle evolution, we compute the growth rate of the pair cascade as a function of the accelerating electric field in the presence of black-body and power-law ambient photon fields. 
Informed by the numerical results, we derive simple analytical expressions for the peak growth rate and the corresponding electric field. We show that for certain parameters, which can be realized in the vicinity of accreting supermassive black holes at the centers of active galactic nuclei, the pair cascade may well be regulated by inverse Compton scattering in the deep Klein-Nishina regime and triplet pair production. We present indicative examples of the escaping gamma-ray radiation from the gap, and discuss our results in application to the TeV observations of radio galaxy M87.
\end{abstract}

\keywords{acceleration of particle -- gamma-rays: galaxies -- galaxies: active, individual (M87) -- radiation mechanisms: non-thermal}


\section{Introduction} \label{sec:intro}
The production of a relativistic electron-positron pair from the absorption of a gamma-ray photon is the starting point of an avalanche of pairs and gamma rays, known as an electromagnetic (EM) cascade. Inverse Compton-pair cascades, in particular, are initiated when gamma-rays are absorbed on a soft photon field to produce relativistic pairs, which in turn inverse Compton scatter off the same soft photons to produce more gamma-rays. 

In the simplest case, where the pairs are not accelerated, only a few generations of them are expected to be produced before the energy of gamma-rays is degraded to a value below the threshold for $\gamma\gamma$ pair creation, thus ceasing the EM cascade. At this point, gamma-rays can escape unimpeded from the region of their production, with their initial energy being shared among the secondary pairs and photons.
As long as the energy density of the secondary photons is lower than that of the background soft photon field, the cascade is considered to be {\sl linear}, i.e., the interactions between secondary pairs and photons are negligible.

Linear inverse Compton-pair cascades in isotropic photon fields\footnote{The development of these cascades in anisotropic photon fields was studied by \cite{protheroeetal92}.} have been discussed in the context of compact X-ray sources \citep{aharonianetal85, mastichiadis86,zdziarski_2009} and applied to the propagation of ultra-high-energy (UHE) gamma-rays in the intergalactic medium (IGM) \citep{gouldrephaeli78,protheroe86, zdziarski_88, coppiaharonian97}. In the latter case, the initial Compton scatterings of UHE pairs on low-energy background photons (e.g., cosmic microwave background) can take place deep in the Klein-Nishina regime (i.e., $s\gg (m_e c^2)^2$, where $s$ is the squared center-of-mass energy of the interaction). In this regime, triplet pair production (TPP; $e\gamma_b \rightarrow ee^{+}e^{-}$) can be an additional source of electron cooling. Although TPP is a third-order quantum electrodynamical process, it can operate in direct competition to inverse Compton scattering (ICS) for $s\gtrsim 100$~MeV$^2$ \citep{haug_81,mastichiadisetal86, mastichiadis_91, anguelov_99}. Because of the very high interaction energies required, the role of TPP has been mainly investigated in the context of UHE-photon induced cascades in the IGM \citep{lee_98, settimo_15, wang_17}.

The vicinity of supermassive black holes in active galactic nuclei (AGN), where strong electric fields may be present and copious soft photons are emitted by the accretion disk, is another astrophysical environment where linear Compton-pair cascades have been studied 
\citep[e.g.,][]{bednarek_95, levinson_11,broderick_2015,hirotanipu16, chen_18,katsoulakosrieger18,levinsoncerutti18}.
In this case, the physics of the cascade are coupled to the acceleration of particles by the component of the electric field parallel to the magnetic field over the length of a charge-starved region, the so-called gap\footnote{There is an analogy to pair cascades in pulsar magnetospheres \citep{rudermansutherland75}, although the physical processes related to the production of gamma-ray photons are different  \citep[see e.g.,][and references therein]{timokhin_2013, timokhin_2019}.}. A seed electron that enters the gap can accelerate to high enough energies to up-scatter disk photons and produce gamma-rays, which in turn pair-produce on the disk photons, thus initiating a pair cascade in the gap. In contrast to the cascades induced in the IGM, the secondary electrons are continuously accelerated inside the gap. This leads to an exponentiation of their number density until they provide the necessary charge density to sort out the electric field and quench acceleration.

In this paper, we study the generic properties of inverse Compton cascades in pair-producing gaps in the Thomson and Klein-Nishina regimes with the inclusion of triplet pair production. More specifically, we compute the growth rate of secondary pairs as a function of the electric field strength for a background soft photon field (with a black-body or power-law energy distribution) and study the effects of triplet pair production on the development of these cascades (e.g., growth rate and maximum particle energy).  This paper is structured as follows. In \sect{model} we qualitatively discuss the general properties of Compton-pair cascades with triplet pair production. In \sect{numerical} we outline the adopted numerical scheme for the calculation of the cascade and present our results on the pair growth rate in \sect{results}. We discuss their astrophysical implications in \sect{astro} and conclude with a discussion on the model's caveats in \sect{discussion}.

\section{General considerations}\label{sec:model}
We consider a stationary rectilinear particle accelerator immersed in an isotropic time-independent ambient photon field with energy density $U_{ph}\equiv \int d\epsilon \, \epsilon n_b(\epsilon)$, where $n_b(\epsilon)$ is the differential photon number density. The accelerator is characterized by a constant electric field with strength $E_0$. The electric field is assumed to be almost parallel to the local magnetic field, which has no curvature. The size of the acceleration region is taken to be much larger than all relevant mean free paths of relativistic pairs, thus allowing us to follow the full development of the pair cascade. 

Within our toy model, we treat electrons and positrons as identical particles with initial Lorentz factor $\gamma_0$. Upon entering the acceleration region, pairs will gain energy from the electric field, and at the same time, lose energy due to ICS (or TPP) on the background soft photon field. Energy losses due to synchrotron and curvature radiation can be neglected  under our assumptions (i.e., the particle pitch angle and the curvature of magnetic field lines are both approximately zero). 

As long as the scatterings occur in the Thomson regime, particles can accelerate from their initial Lorentz factor to a maximum value $\gmax$ where the continuous energy losses balance the energy gains:
\eqb
\label{eq:gmax}
\gmax=\left(\frac{3q_eE_0}{4\sth U_{ph}}\right)^{1/2}.
\eqe
In the above equation $c$ is the speed of light, $\sth$ is the Thomson cross section, and $q_e, m_e$ are the electron charge and mass. The scattering of soft photons with energy  $\xav \equiv \int d\epsilon \, \epsilon n_b(\epsilon)/\int d\epsilon \,n_b(\epsilon)$ by pairs with Lorentz factor $\gmax$ will occur in the Klein-Nishina regime,  if $E_0$ exceeds a critical value that depends solely on the properties of the target photon field, namely:
\eqb 
E_{0}^{\rm (KN)} = \frac{4 \sth U_{ph}}{3 q_e} \left(\frac{m_e c^2}{\xav}\right)^2,
\label{eq:Ekn}
\eqe 
where we used the condition $\gmax \xav \gtrsim (3/4) m_e c^2$ and \eq{gmax}. In particular, for a black-body photon field with temperature $T$ we find $E_0^{\rm (KN)} \propto T^2$.  Using this critical value of the electric field, we can distinguish the following two regimes.

{\it Low-$E_0$ regime:} for $E_0<E_0^{\rm (KN)}$ the acceleration process is balanced by energy losses mostly in the Thomson regime and the particle Lorentz factor saturates at the value given by \eq{gmax}. As a result, the up-scattered photons, of energy $\epsilon_1 \sim \gmax^2 \xav$, are below the threshold for $\gamma\gamma$ production with the average energy soft photons (i.e., $\epsilon_1 \xav < 2(m_e c^2)^2$). Still, pair production is expected to occur at some level due to $\gamma \gamma$ interactions with photons from the high-energy tail of the distribution. We note that TPP is not relevant in this regime due to the much lower interaction rates \citep[see e.g., Fig.~4 in][]{mastichiadis_91}.
The number density of photons at $\epsilon \gg \xav$ together with the rate of acceleration will essentially determine the growth rate of the produced pairs. For a black-body photon field, there is essentially no  free parameter other than $E_0$, since the photon number density at $\epsilon \gg  \langle\epsilon\rangle \simeq 2.7 k_B T$ is just given by the Wien part of the spectrum. However, in the more general case of a power-law photon field, the exact growth rate of the number of pairs is expected to depend on several free parameters: the maximum and minimun cutoff energies ($\epsilon_{\max}$, $\epsilon_{\min}$), the photon index ($\Gamma$), and the energy density ($U_{ph}$). 

{\it High-$E_0$ regime:} for $E_0>E_0^{\rm (KN)}$, pairs can achieve much higher Lorentz factors than $\gmax$ (see \eq{gmax}); the highest energy pairs can up-scatter soft photons in the Klein-Nishina regime, where the energy losses are not strong enough to balance the energy gains by the acceleration. As a result, the maximum particle energy does not saturate at $\gmax$, but continues to increase till another process (e.g., physical escape from the system) settles in. 
Quickly upon entering the acceleration region, the seed pairs can achieve high enough energies to produce  secondary pairs off the photon field either directly through TPP or indirectly through the combination of ICS and $\gamma\gamma$ absorption. In both cases, the produced pairs will also be accelerated by the strong electric field, and their number will exponentiate.
  
In both regimes, a pair cascade will develop in the acceleration region. 
The cascade is, in good approximation, one-dimensional as long as the magnetic deviations of the particle  trajectories over the relevant particle interaction lengths are smaller than the emission cone of the pairs. Our goal is to compute the growth rate of the ensued linear pair cascade as a function of $E_0$ for different ambient photon fields. In the following section, we describe the numerical approach we adopted for the study of the pair cascade. 

\section{Numerical approach}\label{sec:numerical}
The physical problem outlined in \sect{model} can be mathematically formulated using the concept of 
transport equations that describe the evolution of three particle populations, namely the seed pairs that enter the acceleration region, the secondary pairs produced by $\gamma \gamma$ and TPP processes, and the photons emitted by both pair populations via ICS:
\eqb
\label{eq:prim}
{\partial n^{(1)}_{e} \over \partial{t}}+q_e c E_0  \frac{\partial n^{(1)}_{e}}{\partial \gamma}+\mathcal{L}^{ics}_{e}+\mathcal{L}^{tpp}_{e}&=&\mathcal{Q}^{inj}_e  \\ 
{\partial n^{(2)}_{e}\over \partial{t}}+q_e c E_0 \frac{\partial n^{(2)}_{e}}{\partial \gamma}+\mathcal{L}^{ics}_{e}&=&\mathcal{Q}^{tpp}_{e}+\mathcal{Q}^{\gamma\gamma}_{e} \\
\label{eq:sec}
{\partial n_\gamma \over \partial{t}}+\mathcal{L}^{\gamma\gamma}_{\gamma} & = & \mathcal{Q}^{ics}_{\gamma}.
\label{eq:phot}
\eqe
Here, $n^{(1)}_{e}, n^{(2)}_{e}$, and $n_{\gamma}$ denote respectively the differential number densities of seed pairs, secondary pairs, and photons produced by both populations of pairs. The operators $\mathcal{Q}_i^j$ and $\mathcal{L}_i^j$ denote injection and loss terms of the particle species $i$ due to the process $j$. 

To solve the particle transport equations, we employ the numerical code of \cite{mast_95}, but with two main modifications: the implementation of the triplet pair production process and the use of a more accurate scheme for the treatment of electron losses in the Klein-Nishina regime. For completeness, we summarize the different terms appearing in eqs.~(\ref{eq:prim})-(\ref{eq:phot}) briefly.

{\sl Pair injection}. Seed pairs with some low Lorentz factor are injected into the acceleration region at a rate $Q_e^{inj}=Q_0\delta(\gamma -\gamma_0)H(t)$. Here, we adopted (without loss of generality) $\gamma_0=1$. We checked that the derived exponential growth rates (see \sect{results}) do not depend on the choice of $\gamma_0$ or of the injection profile (e.g., continuous versus instantaneous injection). 

{\sl Inverse Compton scattering}. The term  $\mathcal{L}^{ics}_{e}$ takes into account energy losses in both the Thomson and Klein-Nishina regimes according to Eqs.~(5.7) and (5.17) of \cite{blumenthal_70}. For the photon production rate $\mathcal{Q}^{ics}_{\gamma}$ we use Eq.~(2.48) of \cite{blumenthal_70}. 

{\sl $\gamma \gamma$ pair production}.  The pair injection term $\mathcal{Q}^{\gamma\gamma}_{e}$ and photon attenuation rate $\mathcal{L}^{\gamma\gamma}_{\gamma}$ are given respectively by Eqs.~(57) and (54) of \cite{mast_95}. 

{\sl Triplet pair production}. We consider the energy losses of the recoiling electron (or positron) to be continuous. For the single-particle loss rate $q_{tpp}$ we adopt the Eq.~(29) of \cite{mastichiadis_91}, which holds away from the interaction threshold ($\gamma \epsilon >10^3 m_e c^2$). The exact functional form of $q_{tpp}$ near the threshold is not important, since ICS losses dominate in this energy regime \citep{mastichiadis_91}. The loss term appearing in \eq{prim} is modeled as $\mathcal{L}_{e}^{tpp}(\gamma)=\int d\epsilon \, n_b(\epsilon) q_{tpp}(\gamma \epsilon/m_e c^2)$. Following \cite{lee_98}, we model the TPP production rate as $\mathcal{Q}_e^{tpp}(\gamma')=q_0 \gamma^{'-\delta} \int d\gamma \, n^{(1)}_e(\gamma)\int d\epsilon \,  \sigma_{tpp}(\gamma \epsilon/m_e c^2) n_b(\epsilon)$, where $\sigma_{tpp}$ is the total TPP cross section \citep{haug_81, mastichiadis_91}. We adopt the asymptotic value of $7/4$ for the exponent $\delta$ \citep[see also Fig.~2 in][]{mastichiadis_91}, while noting that our results are not sensitive on the exact value of $\delta$. The normalization constant $q_0$ is determined by the requirement of energy conservation, i.e., $\int d\gamma \,  n^{(2)}_e(\gamma) \gamma  \mathcal{Q}_e^{tpp}(\gamma) = \int d\gamma \, n^{(1)}_e(\gamma) \gamma \mathcal{L}_e^{tpp}(\gamma)$.

Finally, we note that the numerical code solves for the combined distribution $n_e=n_e^{(1)}+n_e^{(2)}$. Although the code does not differentiate between seed pairs and secondary pairs, any exponentially growing solutions of $n_e$ are purely determined by the secondaries. This can be easily understood by considering the extreme case of no pair production ($n_e^{(2)}=0$). In this case, one can show by  solving \eq{prim} with $\mathcal{L}^{tpp}_{e}=0$ that the lepton number $\int n_e(\gamma) d\gamma$ increases only linearly with time.

\section{Results}\label{sec:results}
Using the numerical code described in the previous section, we compute the growth rate ($\lambda$) of exponentially increasing solutions for a wide range of electric field strengths $E_0$ and different ambient photon fields. 
More specifically, we perform a linear fit to $y(t)\equiv\ln\left[\int d\gamma \, n_e(\gamma, t)\right]$, since its slope gives directly the rate of exponentially growing solutions. 
In what follows, we compute $\lambda$ for a black-body photon field of temperature $T$ (\sect{BB}) and for a power-law photon field with photon index $\Gamma$ and energy density $U_{ph}$, extending from $\epsilon_{\min}$ to $\epsilon_{\max}$ (\sect{PL}). 

\subsection{Black-body photons}\label{sec:BB}
 \begin{figure}
 \centering 
 \includegraphics[width=0.47\textwidth]{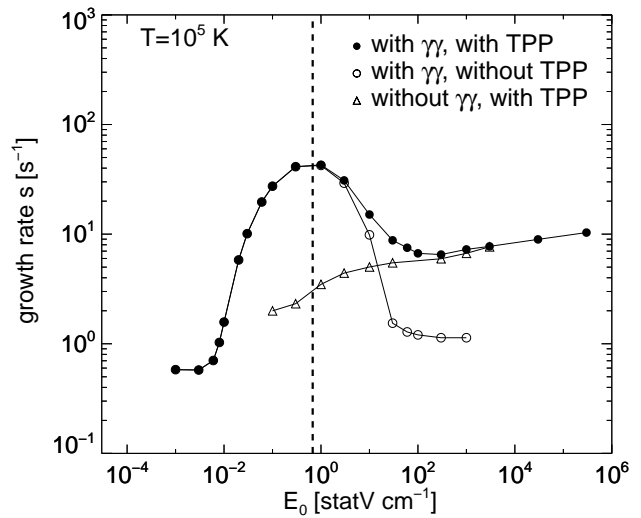} 
 \caption{Growth rate of the number of pairs as a function of the electric field strength $E_0$, for a black-body photon field with temperature $T=10^5$~K. Different symbols are used to plot the growth rates when different processes are taken into account (for details, see inset legend). The vertical dotted line indicates the critical value of the electric field given by \eq{Ekn}.}
 \label{fig:s-BB}
\end{figure}

\begin{figure*}
    \centering
 \includegraphics[width=0.32\textwidth]{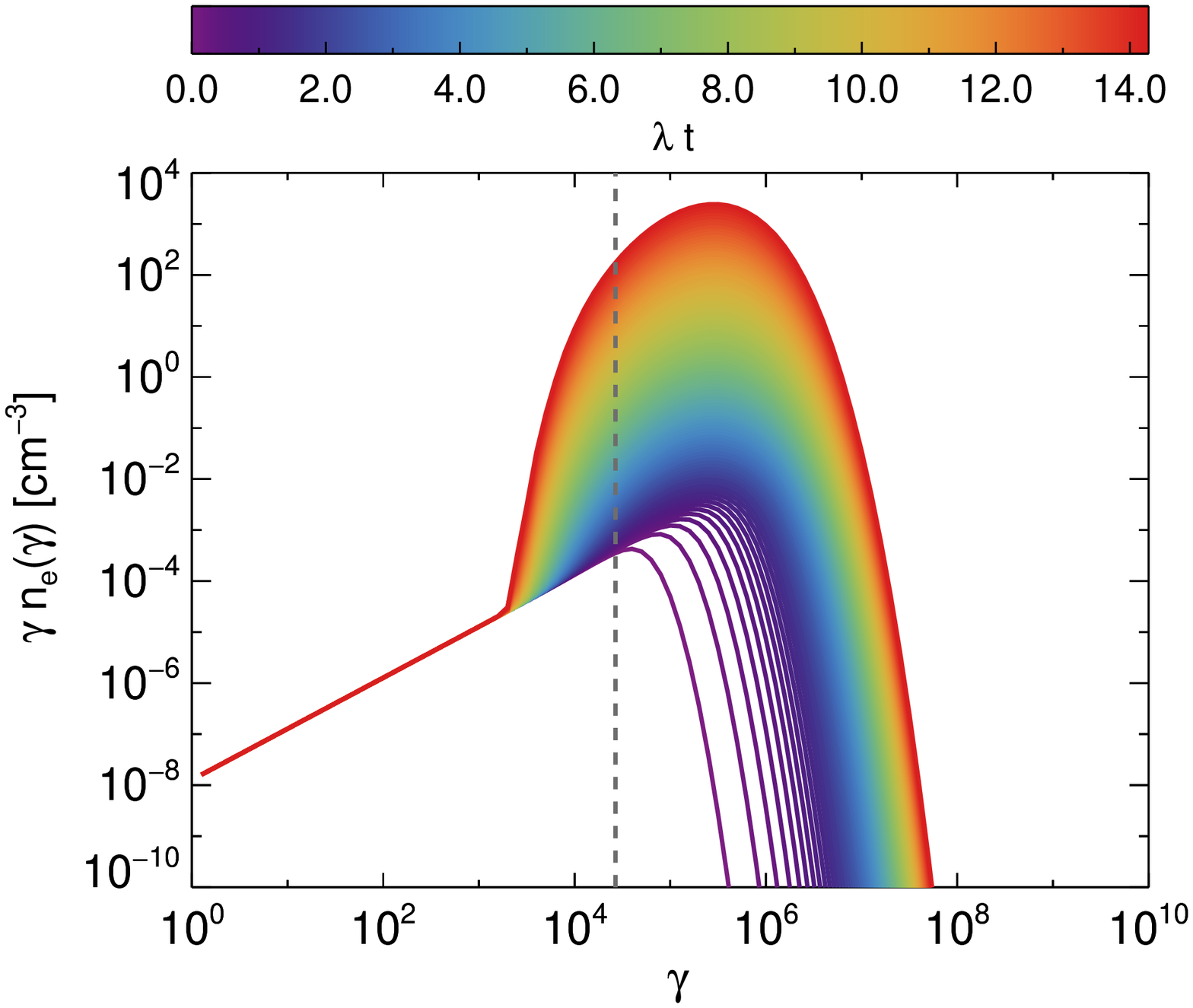}
 \includegraphics[width=0.32\textwidth]{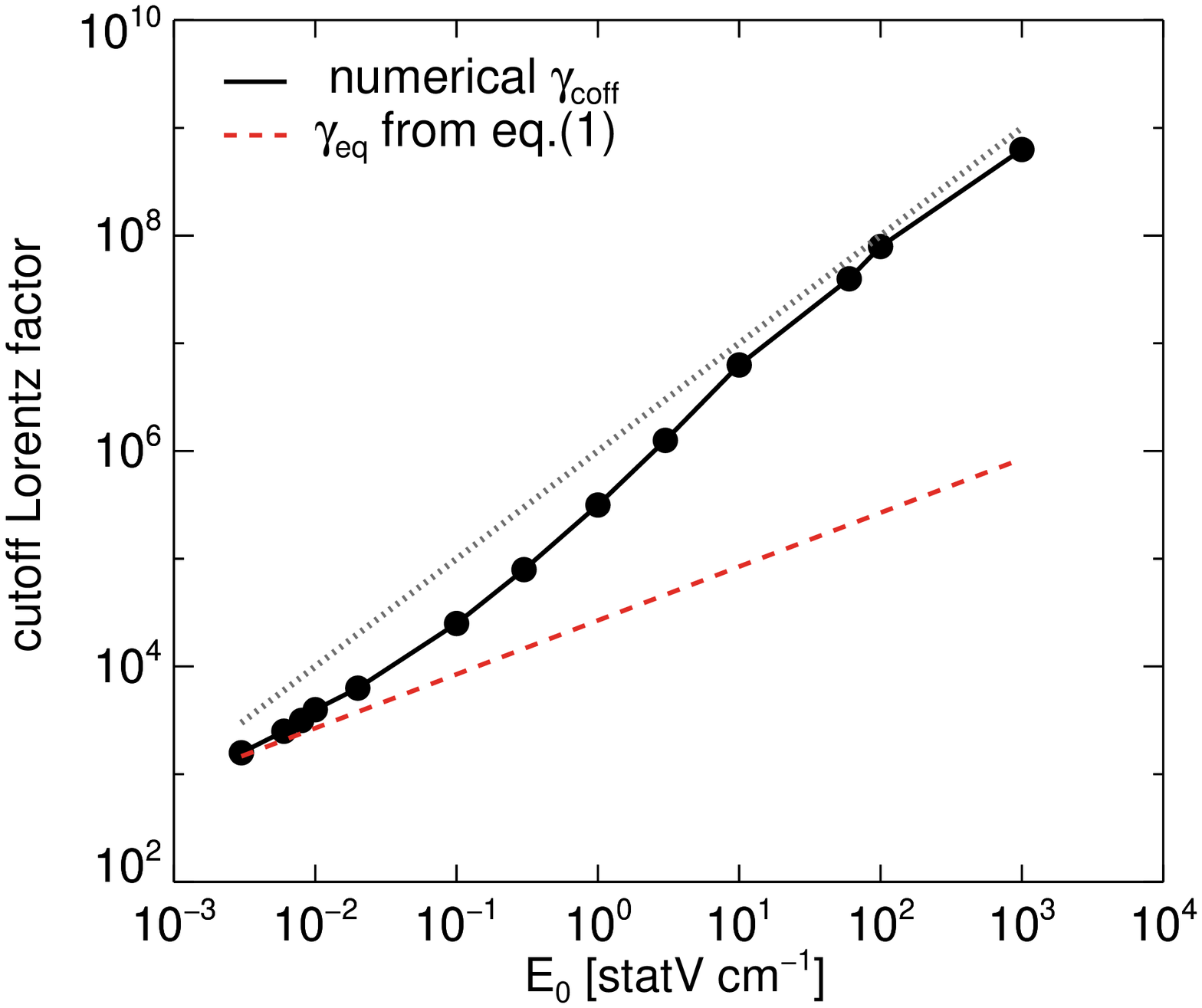} 
 \includegraphics[width=0.32\textwidth]{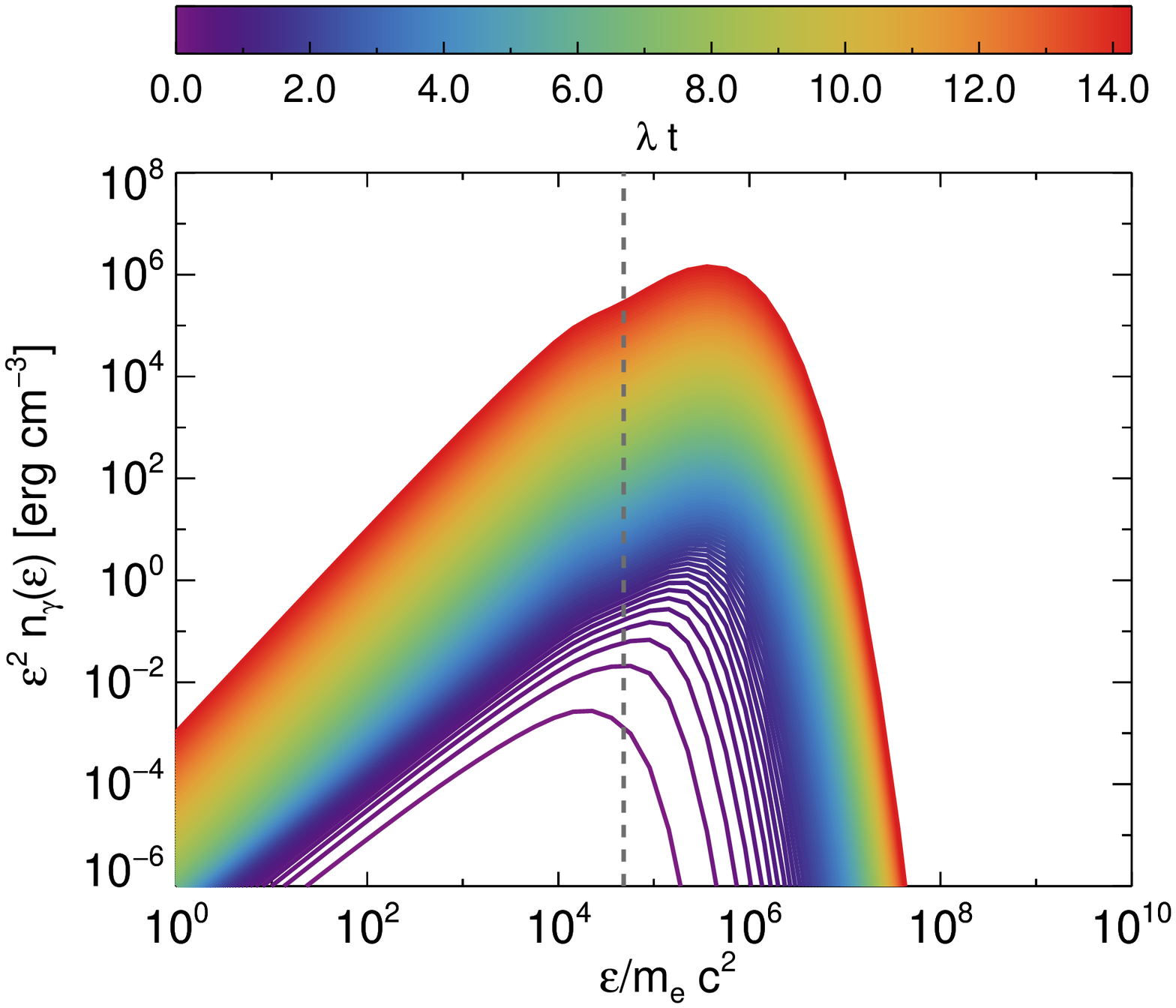}
    \caption{{\sl Left panel:} Temporal evolution of the pair energy distribution (colored lines) for the same photon field used in Fig.~\ref{fig:s-BB} and $E_0=1$~statV cm$^{-1}\simeq E_0^{\rm (KN)}$. Here, both $\gamma\gamma$ and triplet pair production processes are taken into account. The color bar indicates the time (in units of the growth time $\lambda^{-1}$). The dashed vertical line marks the saturation Lorentz factor for IC losses in the Thomson regime given by \eq{gmax}. {\sl Middle panel:} Cutoff Lorentz factor of the pair energy distribution computed numerically for different electric field strengths (black symbols). The dashed red line shows $\gmax$ defined in \eq{gmax}. A line of slope one (dotted grey) is overplotted to illustrate the asymptotic scaling of the cutoff Lorentz factor on $E_0$, when TPP dominates. {\sl Right panel:} Snapshots of the IC photon energy spectrum (colored lines) produced in the gap. The dashed vertical line marks the typical up-scattered photon energy of $4 \,k_B T \gmax^2 $. The color bar indicates the time (in units of the growth time $\lambda^{-1}$) or, equivalently, the location within the gap at which the spectrum is produced. The last snapshot shown here should be considered as the escaping photon spectrum from a gap with size $h=14 \, c/\lambda \simeq 10^{10}$~cm.}
    \label{fig:elec-BB}
\end{figure*}

\fign{s-BB} presents the exponential growth rate $\lambda$ computed numerically for different electric field strengths $E_0$ when: (i) both $\gamma \gamma$ and triplet pair production processes are taken into account (filled symbols), (ii) only triplet pair production is taken into account (open triangles), and (iii) only $\gamma \gamma$ pair production is included (open circles). Note that the growth rate in the absence of pair production asymptotes to a constant value  instead of approaching zero (see e.g., filled symbols at $E_0 \lesssim 10^{-2}$~statV cm$^{-1}$ and open circles at $E_0 >10^3$~statV cm$^{-1}$). This is a result of our choice for the injection of seed particles (see also \sect{numerical}). With a continuous injection,  even in the absence of any pair production processes ($n_e^{(2)}=0$), the number of seeds is expected to increase linearly with time. These non-exponential growing solutions are reflected to the non-zero values of the numerically derived growth rates\footnote{We verified that $\lambda \rightarrow 0$ for sufficiently low or high (in the absence of TPP) electric field strengths, if seed particles are instantaneously injected into the gap}.

The actual shape of the curve $\lambda(E_0)$ is determined by the convolution of the ambient photon field energy distribution with the energy-dependent cross sections of the relevant physical processes. In the low-$E_0$ regime (i.e., for $E_0 < E_0^{\rm (KN)}$), the growth rate is determined essentially by the $\gamma \gamma$ pair production process. However, in the high-$E_0$ regime ($E_0 > E_0^{\rm (KN)}$), where the $\gamma \gamma$ cross section decreases (see open circles), triplet pair production begins to take over and eventually dominates the total growth rate of pairs at $\gtrsim 100 \, E_0^{\rm (KN)}$ (open triangles). In this regime, the slow increase of $\lambda$ with increasing $E_0$ reflects the logarithmic dependence of the TPP cross section (see also \fign{s-PL} and text in \sect{PL}, for more details).

As an indicative example, we show in \fign{elec-BB} (left panel) the temporal evolution of the pair energy distribution computed for $E_0=1$~staV cm$^{-1} \simeq E_0^{\rm (KN)}$, when both pair production processes are taken into account. The dashed vertical line marks the saturation Lorentz factor assuming ICS losses in the Thomson regime (see \eq{gmax}). The energy spectrum can be modeled as $\gamma n_e(\gamma) \propto \gamma e^{-\gamma/\gamma_{\rm coff}}$, where $\gamma_{\rm coff}$ is generally time-dependent. Before the number of pairs exponentiates (i.e., $t\lesssim \lambda^{-1}$), we find that $\gamma_{\rm coff}\propto t$ and $\gamma_{\rm coff} \gtrsim \gmax$. The extension of the spectrum beyond $\gmax$ suggests that the up-scattering of black-body photons with energy $\langle \epsilon\rangle=2.7 k_BT$ by pairs with $\gamma\gtrsim \gmax$ takes place in the Klein-Nishina regime, already from $E_0\sim E_0^{\rm (KN)}$. 

Soon after the onset of the cascade (i.e., $t\gtrsim \lambda^{-1}$), the cutoff Lorentz factor of the distribution freezes at a value that depends on $E_0$. This is illustrated in the middle panel of \fign{elec-BB}, where we plot the time-independent $\gamma_{\rm coff}$ determined numerically for different electric field strengths. The dashed red line denotes the saturation Lorentz factor considering particle losses only in the Thomson regime (see \eq{gmax}). As soon as pair production becomes important, we find that $\gamma_{\rm coff}> \gmax$. For the adopted parameters, this occurs at $E_0 \gtrsim 10^{-2}$~statV cm$^{-1}$ (see also Fig.~\ref{fig:s-BB}).
In the high-$E_0$ regime, in particular, the dependence of $\gamma_{\rm coff}$ on $E_0$ can be derived from the following condition: any pairs injected via the TPP process should reach the Lorentz factor $\gamma$ of the parent electron within the mean free path of the process, namely $m_e \gamma c^2 \approx e E_0 \ell_{tpp}(\gamma)$ with $\ell^{-1}_{tpp}(\gamma) \approx \int d{\epsilon} \,  n_b(\epsilon) \sigma_{tpp}(\gamma \epsilon/m_e c^2)$. For a black-body photon distribution, we find $\ell_{tpp}(\gamma) \approx$ constant and $\gamma_{\rm coff}^{(tpp)} \propto E_0$, in agreement with the numerical results for $E_0\gtrsim 30$~statV cm$^{-1}$, where the pair production is regulated via the TPP process (see also \fign{s-BB}).

The right panel of \fign{elec-BB} shows snapshots of the photon energy spectra produced by ICS off an ambient black-body photon field by the pairs in the accelerating gap (see left panel of the same figure). For reference, the typical photon energy of $4 \gmax^2 k_B T$ is also indicated (dashed grey line). Each snapshot corresponds to a certain time (in units of $\lambda^{-1}$) or, equivalently, a certain location within the gap. If the physical size of the latter is taken to be $h=14 \, c/\lambda=10^{10}$~cm, then the last snapshot shown in the figure would correspond to the spectrum of photons escaping from the gap. Note that this would not necessarily be  the observed spectrum, because other processes operating outside the gap could still alter it (for further discussion, see \sect{m87}).

\subsection{Power-law photons}\label{sec:PL}
\begin{figure*}
 \centering 
 \includegraphics[width=0.329\textwidth, trim=33 0 0 0]{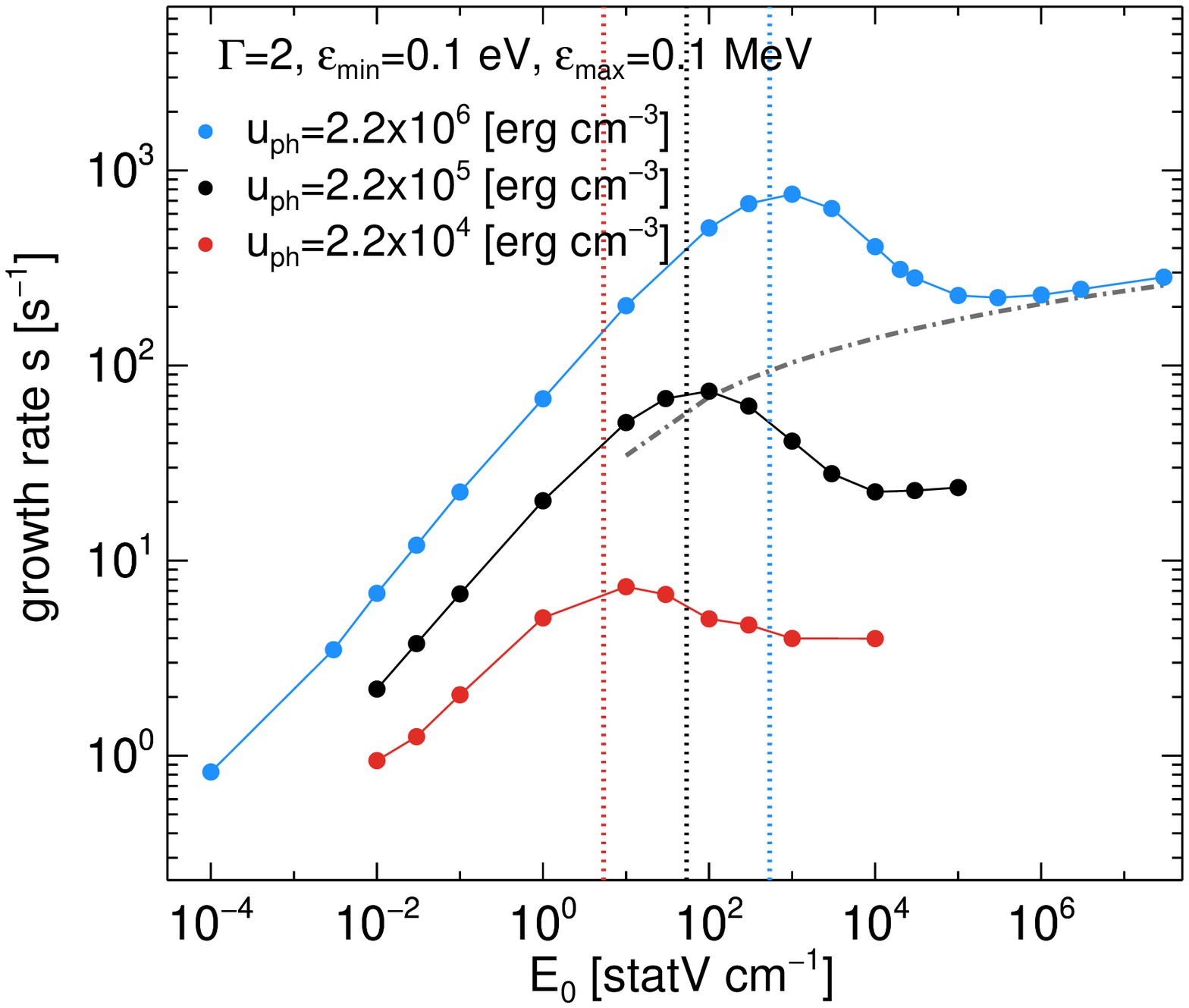} 
 \includegraphics[width=0.329\textwidth,trim=33 0 0 0]{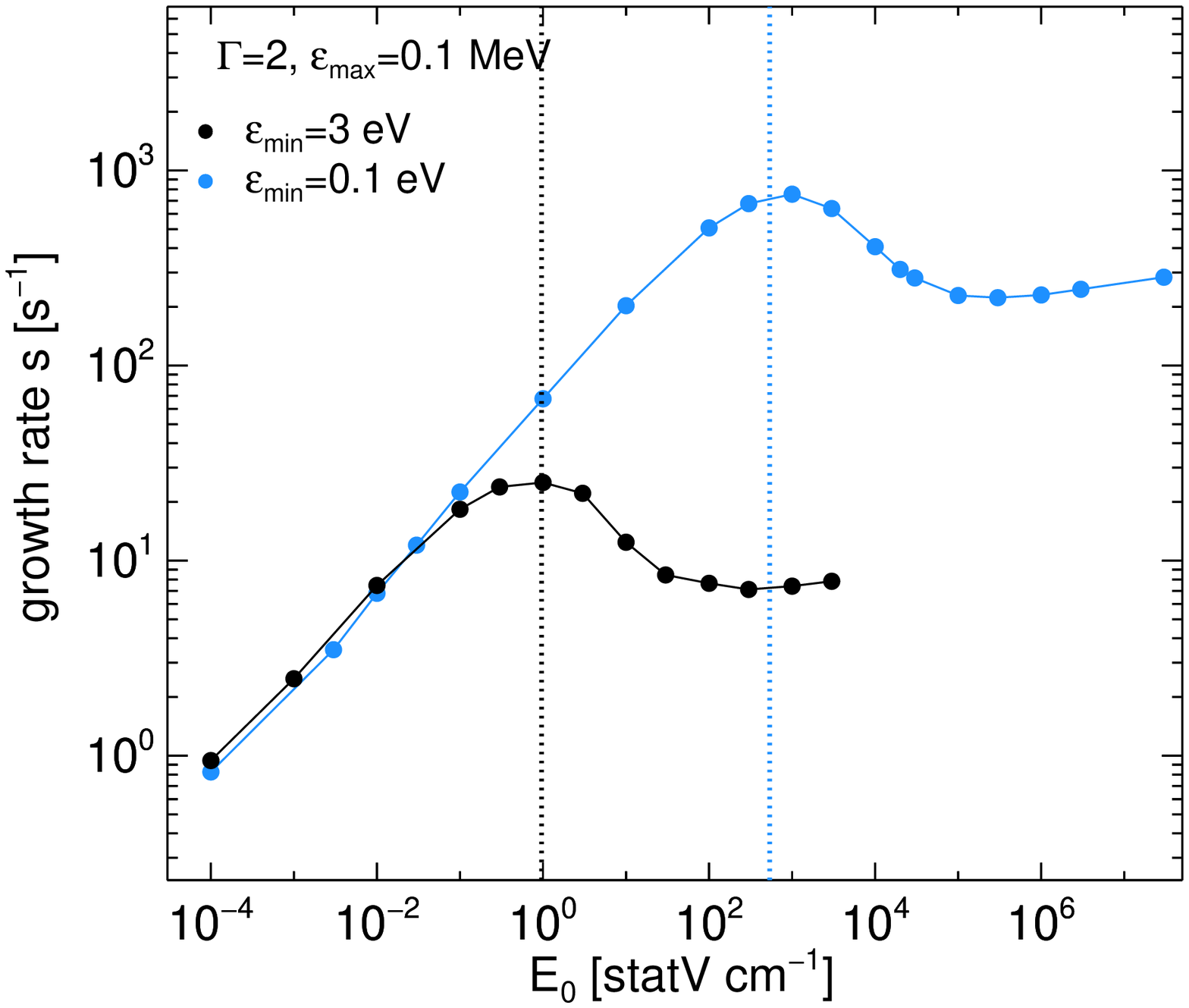}
 \includegraphics[width=0.329\textwidth,trim=33 0 0 0]{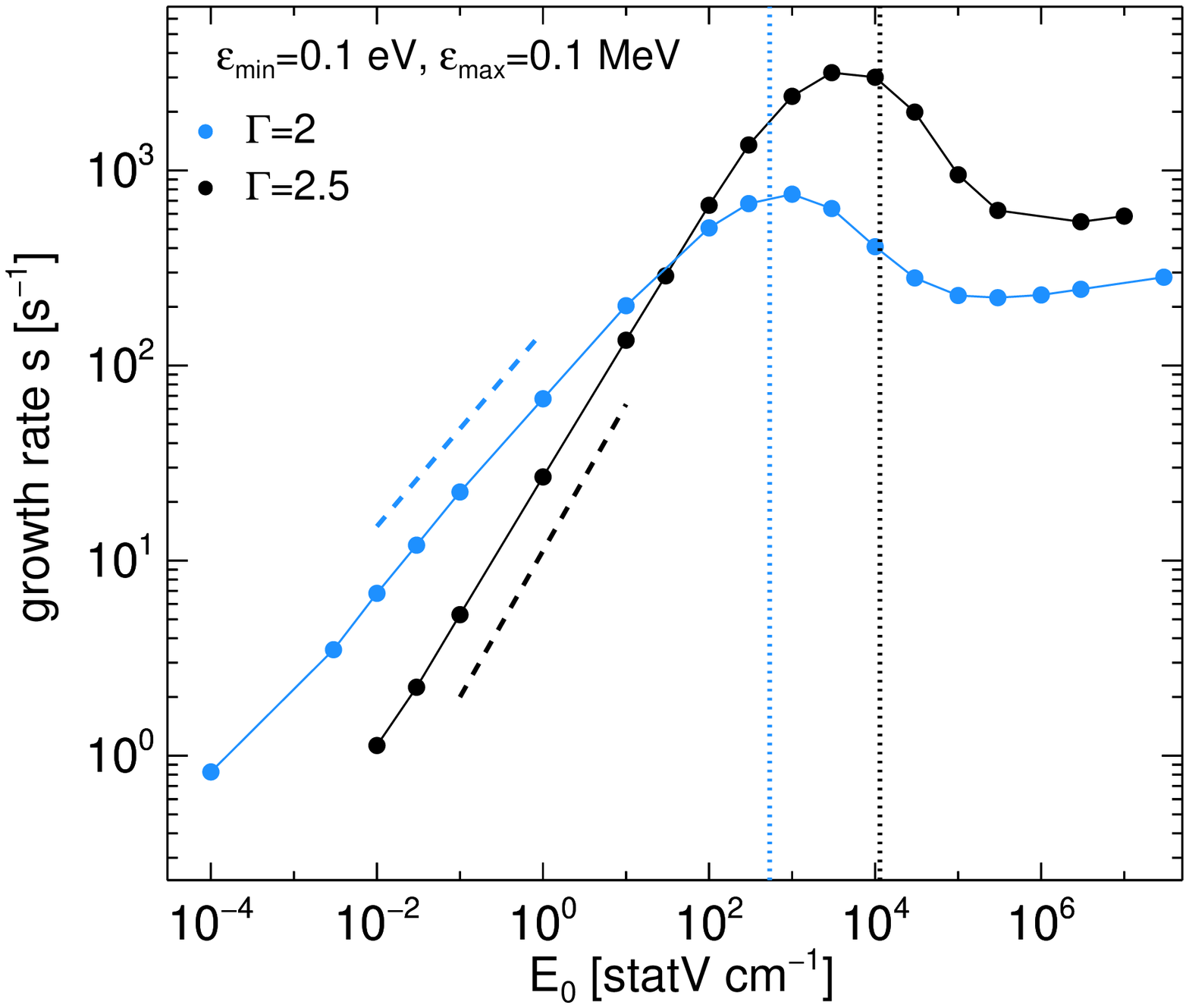}
  \caption{Same as in Fig.~\ref{fig:s-BB}, but for a power-law target photon field with $\Gamma=2$, $\epsilon_{\min}=0.1$~eV, $\epsilon_{\max}=0.1$~MeV, and $U_{ph}=2.2\times10^6$~erg cm$^{-3}$, unless stated otherwise. We show the dependence of the growth rate (from {\sl left} to {\sl right}) on the photon energy density, minimum photon energy, and photon index. In all panels, vertical dotted lines mark the critical value of $E_0$ given by \eq{Ekn}. The dashed lines plotted in the right panel have slopes  $(\Gamma-1)/2$, as analytically predicted (see \eq{kappa-E0}). We also predict $\lambda\propto \ln E_0$ for $E_0 \gg E_0^{(\rm KN)}$ (see \eq{s-tpp}), as shown in the left panel (dash-dotted grey line). A colored version of this plot is available online.}
 \label{fig:s-PL}
\end{figure*}
\fign{s-PL} presents the exponential growth rate $\lambda$ as a function of $E_0$ for an ambient photon field with a power-law energy distribution. Here, both $\gamma \gamma$ and triplet pair production are considered. Different panels (from left to right) demonstrate the effects of $U_{ph}, \epsilon_{\min}$, and $\Gamma$  on the growth rate (for details, see figure caption and inset legends). By varying the photon energy density, while keeping all other parameters fixed (left panel), we obtain a self-similar family of curves for $\lambda(E_0)$. For example, if we were to shift the black curve to the right and upwards by a factor of $10$ in each direction, we would obtain the blue curve. More specifically, the peak growth rate $\lambda_{\max}$ and the electric field strength at the peak scale linearly with $U_{ph}$. The latter scaling agrees with the predicted one for $E_0^{(\rm KN)}$ (see \eq{Ekn}). 

The growth rate depends also on $\epsilon_{\min}$, as shown in the middle panel, for fixed $U_{ph}$ and $\Gamma$. In particular, we find larger growth rates for lower $\epsilon_{\min}$, as the latter translates to a higher number density of low energy photons, acting as targets for ICS and pair production. Were the black curve shifted to the right by a factor of $\sim 30^2$ and upwards by a factor of $\sim 30$, it would match the blue curve. Thus, for fixed $U_{ph}$ and $\Gamma$ we find  $\lambda_{\max}\propto \epsilon_{\min}^{-1}$. This also translates to $\lambda_{\max}\propto \langle \epsilon\rangle^{-1}$, where the mean photon energy for a power-law photon field is:
\eqb 
\label{eq:mean}
\langle \epsilon \rangle = \left \{
\begin{array}{ll}
\epsilon_{\min} \ln\left(\frac{\epsilon_{\max}}{\epsilon_{\min}}\right)\left(1-\frac{\epsilon_{\min}}{\epsilon_{\max}}\right)^{-1}, &\Gamma=2  \\ \\
\frac{-\Gamma+1}{-\Gamma+2}\frac{\epsilon_{\max}^{-\Gamma+2}-\epsilon_{\min}^{-\Gamma+2}}{\epsilon_{\max}^{-\Gamma+1}-\epsilon_{\min}^{-\Gamma+1}}, & \Gamma \ne 2 
\end{array}
\right. 
\eqe  

Finally, the effects of the photon index
on the growth rate are presented in the right panel of \fign{s-PL}. We find that changes on $\Gamma$  are imprinted on both the peak growth rate and the slope of the power-law segment of $\lambda(E_0)$ at $E_0 < E_0^{\rm (KN)}$. By changing the photon index and keeping all other parameters fixed,  we are effectively varying the mean photon energy of the power-law distribution. This explains why the peak electric field $\approx E_{0}^{\rm (KN)}$ scales as $\langle \epsilon\rangle^{-2}$ (see \eq{Ekn}). We also verified that the peak growth rates of the black and blue curves match, if the black curve is shifted downwards by a factor equal to the ratio of the mean photon energies. 
Our numerical results also suggest that $\lambda\propto E_0^{(\Gamma-1)/2}$ in the low-$E_0$ regime. In what follows, we present analytical arguments to explain the power-law dependence of the growth rate on the photon index.  

The growth rate is proportional to the number of generations of pairs produced per unit time. The latter, denoted as $\kappa$, can be  estimated as $\kappa\approx c/(\ell_{\rm acc}+\ell_{\rm IC}+\ell_{\gamma\gamma})$, where $\ell_{\rm acc}$ is the acceleration length, $\ell_{\rm IC}$ is the IC scattering mean free path, and $\ell_{\gamma\gamma}$ is the $\gamma\gamma$ collision mean free path. When ICS  is in Thomson regime, particles reach a terminal Lorentz factor as determined from \eq{gmax}. 
As these particles IC scatter soft photons from different segments of the power-law energy spectrum, their mean free paths will also be different.
Let $n_b(\epsilon)=n_0(\epsilon/\epsilon_{\min})^{-\Gamma}$ be the differential number density of the ambient photon field. Then, the mean free path for ICS on photons of energy $\epsilon$ is roughly:
\begin{equation}
    \ell_{\rm IC}(\epsilon)\approx \frac{1}{\sigma_Tn_b(\epsilon)\epsilon}=\frac{\epsilon^{\Gamma-1}}{\sigma_T n_0\epsilon_{\min}^{\Gamma}}.
\end{equation}
The ICS on these soft photons produces high energy photons of energy $\epsilon_1\approx\gmax^2\epsilon$, which can then interact with soft photons to produce pairs. The most likely target soft photons for $\gamma \gamma$ pair production are those near the threshold of the interaction, namely their energy is $\epsilon_t\approx (m_ec^2)^2/\epsilon_1$, and the $\gamma\gamma$ mean free path is:
\begin{equation}
    \ell_{\gamma\gamma}(\epsilon_t)\sim \frac{1}{\sigma_{\gamma\gamma}n_b(\epsilon_t)\epsilon_t}=\frac{\epsilon_t^{\Gamma-1}}{\sigma_{\gamma\gamma}n_0\epsilon_{\min}^{\Gamma}}.
\end{equation}
The peak cross section of the  $\gamma\gamma$ pair production process (near the threshold) is similar to the Thomson cross section \citep[$\sigma_{\gamma\gamma}$$\approx\sigma_{T}/5$, ][]{Gould1967}. So, if the target photons for ICS have low $\epsilon$, then $\ell_{\rm IC}(\epsilon)$ is small, but $\ell_{\gamma\gamma}(\epsilon_t)$ of the most-likely-to-be attenuated photons is large, and {\sl vice versa}. As a result, pair production is most efficient on target photons that have similar mean free paths for ICS and $\gamma \gamma$ pair production. This essentially means that $\epsilon\approx\epsilon_t\approx m_ec^2/\gmax$. 

Moreover, in the low-$E_0$ regime, the acceleration length $\ell_{\rm acc}$ is typically much smaller than the interaction mean free paths. The condition $\ell_{\rm acc}(\gmax)\lesssim\ell_{\rm IC}(\epsilon=m_ec^2/\gmax)$ can be written as:
\eqb 
\gmax \epsilon_{\min} \lesssim  m_ec^2\left(\frac{4/3}{\Gamma-2}\right)^{1/(\Gamma-2)}, \Gamma >2 \\
\ln\left(\frac{\epsilon_{\max}}{\epsilon_{\min}}\right) \gtrsim  3/4,  \Gamma=2.
\eqe
Noting that $\gmax \epsilon_{\min} < \gmax \langle \epsilon\rangle < m_e c^2$, both conditions are easily satisfied in the low-$E_0$ regime. Thus, the generation rate of pairs can be estimated by:
\begin{equation}
\label{eq:kappa-E0}
    \kappa\propto \frac{c}{\ell_{\rm IC}(\epsilon=m_ec^2/\gmax)}\propto \gmax^{\Gamma-1}\propto E_0^{\frac{\Gamma-1}{2}}, 
\end{equation}
which roughly agrees with the slope we obtain from the numerical calculations (see right panel in \fign{s-PL}). 

In the high-$E_0$ regime, and in particular at $E_0 \gg E_0^{\rm (KN)}$, pairs are injected into the acceleration region via the TPP process. The mean free path of a high-energy electron with Lorentz factor $\gamma < 2 m_e c^2/\epsilon_{\min}$
is $\propto \gamma^{-\Gamma+1}$, while it has only a logarithmic dependence on $\gamma$ otherwise, i.e., $\ell_{tpp}(\gamma)\propto \ln^{-1}\left(2\gamma \epsilon_{\min}/m_e c^2\right)$. The latter case is typically satisfied in the high-$E_0$ regime, where pairs are ultra-relativistic and their acceleration length is not anymore negligible. Pair production is most efficient when $\ell_{\rm acc}(\gamma)\approx \ell_{tpp}(\gamma)\Rightarrow \gamma \propto E_0$, where the latter relation is derived after neglecting the logarithmic dependence of the TPP mean free path on the Lorentz factor. This condition also gives an estimate of the cutoff Lorentz factor of the pair distribution\footnote{When TPP is the dominant source of secondary pairs, the cutoff Lorentz factor of the pair distribution can be derived by $\ell_{\rm acc}(\gamma_{\rm coff}) \approx \ell_{tpp}(\gamma_{\rm coff})$. 
Thus, $\gamma_{\rm coff}^{(tpp)} \propto E_0^{1/\Gamma}<2 m_e c^2/\epsilon_{\min}$ and  $\gamma_{\rm coff}^{(tpp)}\propto E_0>2 m_e c^2/\epsilon_{\min}$. Typically, only the latter branch is relevant, since at low Lorentz factors pair production is regulated by ICS and $\gamma \gamma$ processes.}, as discussed in Sec.~\ref{sec:BB} for a black-body photon field (see also Fig.~\ref{fig:elec-BB}). The number of pairs in the deep high-$E_0$ regime is therefore estimated as:
\eqb 
\kappa \propto \frac{c}{\ell_{tpp}(\gamma=\gamma_{\rm coff}^{(tpp)})}\propto \ln E_0,
\label{eq:s-tpp}
\eqe 
which agrees with the logarithmic scaling of the growth rate determined numerically (see dash-dotted line in the left panel of \fign{s-PL}).

Based on the self-similarity of the numerical results shown in \fign{s-PL} and the analytical arguments for the dependence of the growth rate on the photon index in the low-$E_0$ regime, we may write the following analytical expression for the growth rate:
\eqb 
\label{eq:s-low}
\lambda \approx \lambda_{\max} \left(\frac{E_0}{E_0^{\rm (KN)}}\right)^{\frac{\Gamma-1}{2}}, \quad E \le E_0^{\rm (KN)}
\eqe 
and its maximum value:
\eqb
\label{eq:smax}
\lambda_{\max} \approx 481 \, \frac{U_{ph, 6}}{\langle\epsilon\rangle_0} \, [{\rm s}^{-1}],
\eqe 
where $U_{ph}\equiv10^6 \, U_{ph,6}$~erg cm$^{-1}$ and $\langle\epsilon\rangle\equiv 10^0 \,\langle\epsilon\rangle_0$~eV.

\begin{figure*}
 \centering 
 \includegraphics[width=0.48\textwidth]{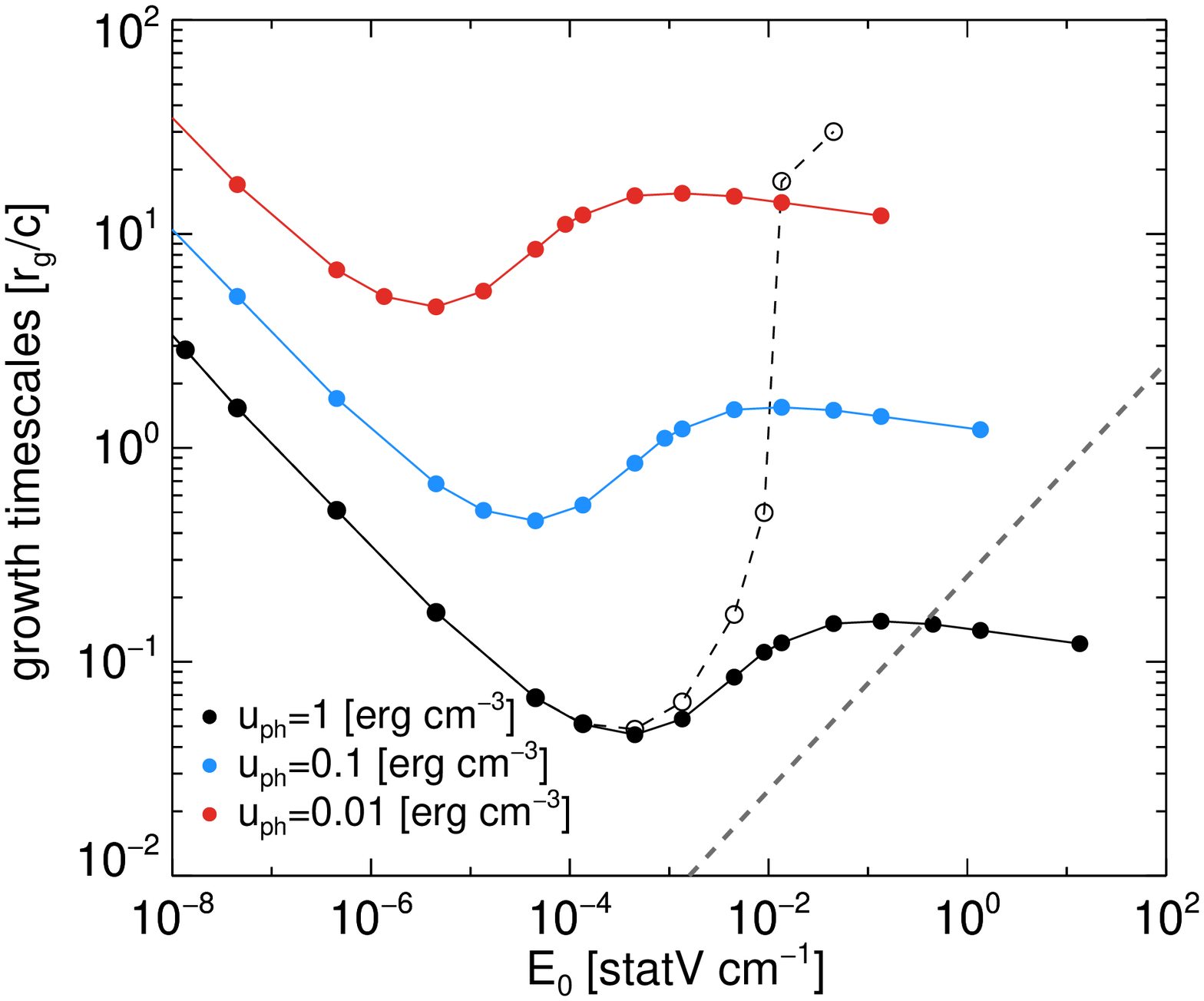} 
  \includegraphics[width=0.48\textwidth]{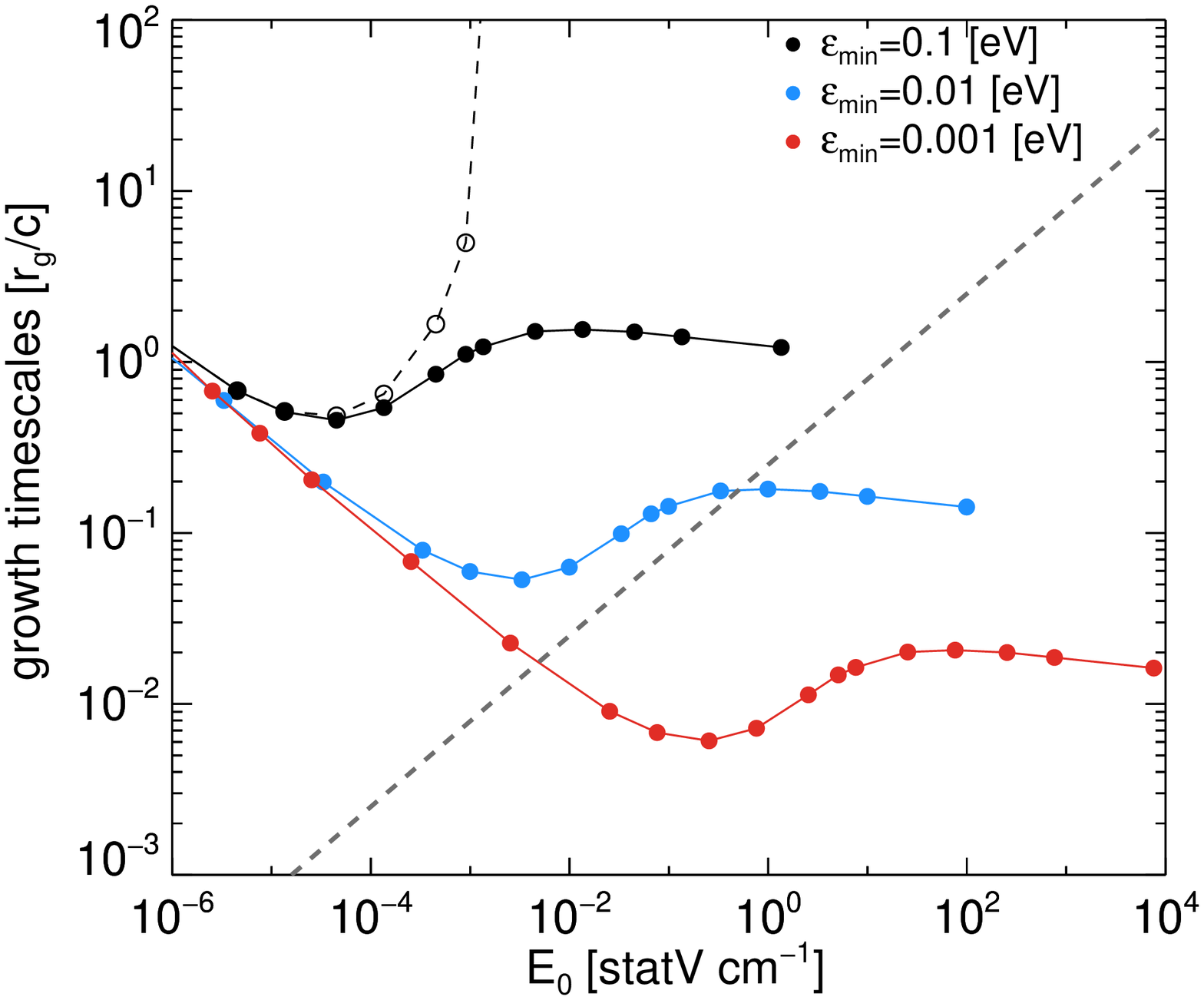} 
 \caption{{\sl Left panel:} Growth timescale  of the number of pairs as a function of the electric field $E_0$ for power-law photon fields with $\Gamma=2$, $\epsilon_{\min}=0.1$~eV, $\epsilon_{\max}=0.1$~MeV, and different energy densities marked on the plot. All curves are obtained by rescaling the numerical results shown in \fign{s-PL} for $U_{ph}=2.2\times10^6$ erg~cm$^{-1}$ according to eqs.~(\ref{eq:Ekn}), (\ref{eq:mean}), and (\ref{eq:smax}). For one case, we also show for illustrative purposes the exponential growth time without TPP (open black symbols). {\sl Right panel:} Same as in the left panel, but for different values of $\epsilon_{\min}$ marked on the plot. Here, $U_{ph}=0.1$~erg cm$^{-3}$.  In both panels, we plot a proxy for the electric field growth time given by \eq{tE} for $B_0=10^2$~G (dashed grey line). All timescales are normalized to $r_g/c$, where $r_g\simeq 2\times10^{15}$~cm, which is also the gravitational radius of M87* \citep{EHT2019}. A colored version of this plot is available online. }
 \label{fig:times}
\end{figure*} 

\section{Astrophysical applications}\label{sec:astro}
A funnel-like region forming at the interface of a jet with the accretion flow of a black hole \citep[e.g.,][]{bednarek_95} or a magnetospheric gap \citep[e.g.,][]{1977MNRAS.179..433B,1992SvA....36..642B,1998ApJ...497..563H} could be considered astrophysical analogues of our simplified particle accelerator.

In the case of a gap in the black-hole magnetosphere, it has been shown that the development of a non-ideal electric field and its screening can be highly time variable \citep{chen_18}. Let us consider a flux tube threading the black hole horizon. At small distances within the tube particles fall into the black hole, while at large distances particles are flung outward by magneto-centrifugal effects. As a result, charge carriers get gradually depleted in the flux tube, leading to the growth of a non-ideal electric field when particles are not enough to carry the necessary current. Before any significant pair production happens, the electric field increases almost quadratically with time, so the time needed for the electric field to grow up to $E_0$ can be approximated as \citep{chen_18}:
\eqb 
\label{eq:tE}
t_E \approx \frac{E_0}{\dot{E}} \approx \left(2\pi\frac{E_0}{B_0}\right)^{1/2} \frac{r_g}{c},
\eqe 
where $r_g=2GM_{\rm BH}/c^2$  
is the gravitational radius of a black hole with mass $M_{\rm BH}$ and $B_0$ is the strength of the poloidal magnetic field near the black-hole horizon. If the acceleration time of the particles is much shorter than the timescale for electric field growth, then the assumption of a constant electric field adopted here is  valid (see \sect{discussion} for a discussion on the validity of this assumption).
A crude criterion for the screening of the gap is that the pair exponentiation timescale becomes  comparable to or shorter than the timescale for the electric field growth given by \eq{tE}.

In \fign{times} we show the growth timescale of the Compton-pair cascade (colored curves) for a  power-law background photon field with different energy densities and minimum energies (for details, see plot legends). 
The cascade timescale is large at small $E_0$, but becomes smaller than $t_E$ (dashed grey line) for sufficiently large $E_0$. 
The equality of the two timescales gives an estimation of the maximum electric field that can develop in an electrostatic gap before it is screened, i.e., before the number density of pairs exceeds the Goldreich-Julian density $n_{\rm GJ}=\Omega B_0 / 2 \pi e c \approx \alpha B_0/ 4 \pi e r_g \simeq 1.7 \times 10^{-5} \, B_{0,2} \,  r^{-1}_{g,15}$~cm$^{-3}$, where $\alpha \sim 1$ is the dimensionless spin of the black hole and $r_{g,15}\equiv r_g/10^{15} \, {\rm cm}$. For certain combinations of $\langle \epsilon\rangle$ and $U_{\rm ph}$, the screening of the electric field  can happen in the regime where the pair cascade is governed by the TPP process and Compton scattering in the deep Klein-Nishina regime (see e.g., black and blue curves in the left and right panels of \fign{times}, respectively):
\eqb 
\label{eq:smax-tE}
\langle\epsilon\rangle_{0}^2 \, U_{ph,0}^{-3/2} \gtrsim 0.02 \, M_9 \, B_{0,2}^{-1/2},
\eqe 
where we used the condition $\lambda_{\max}^{-1}>t_E$, Equations~(\ref{eq:Ekn}),(\ref{eq:smax})-(\ref{eq:tE}), and $M_9\equiv M_{\rm BH}/10^9 M_{\odot}$. 

The characteristic timescale $t_*\equiv t_{E}=\lambda^{-1}$ can also be used to estimate the typical size of the gap as $h=c\,t_*$, which lies in the range $(0.01-1)\,  r_g$ for all the cases displayed in \fign{times}. The maximum electric field that can be developed in the gap dictates also the amount of dissipation in the magnetosphere. The gap power can be estimated as $P_{\rm gap}\sim E_0 J A h$, where $A=\pi r_g^2$ is the cross sectional area of the jet funnel, and $J\sim B c/(4\pi r_g)$ is the Goldreich-Julian current density required by the jet. As a comparison, the jet power is roughly $P_{\rm jet}\sim \alpha B_0^2 A c /(64 \pi)$, and one can write:
\eqb
P_{\rm gap} \sim \frac{16E_0}{B_0} \frac{h}{r_g} P_{\rm jet}. 
\label{eq:Pgap}
\eqe
\begin{figure*}
    \centering
    \includegraphics[width=0.47\textwidth]{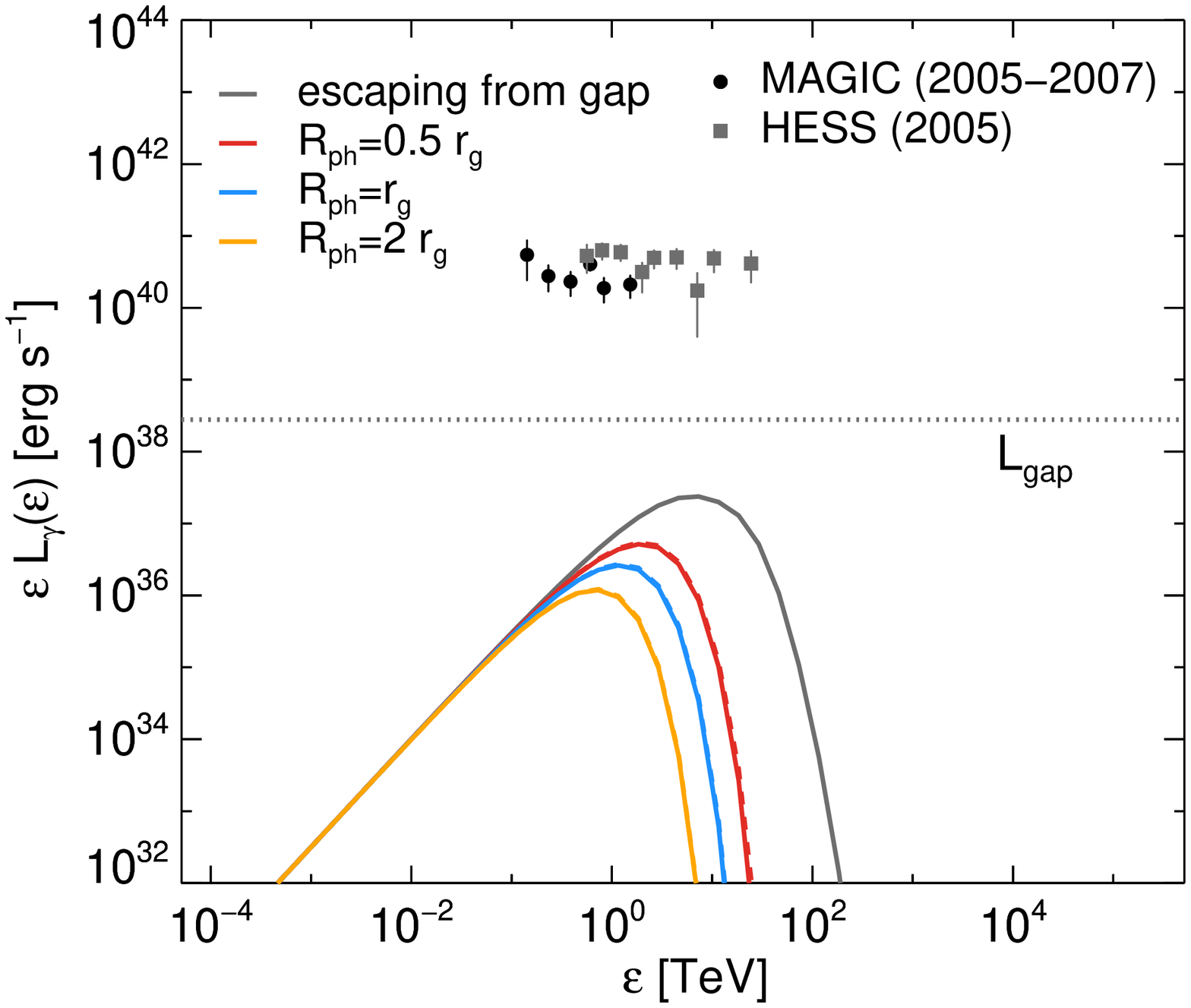}
     \includegraphics[width=0.47\textwidth]{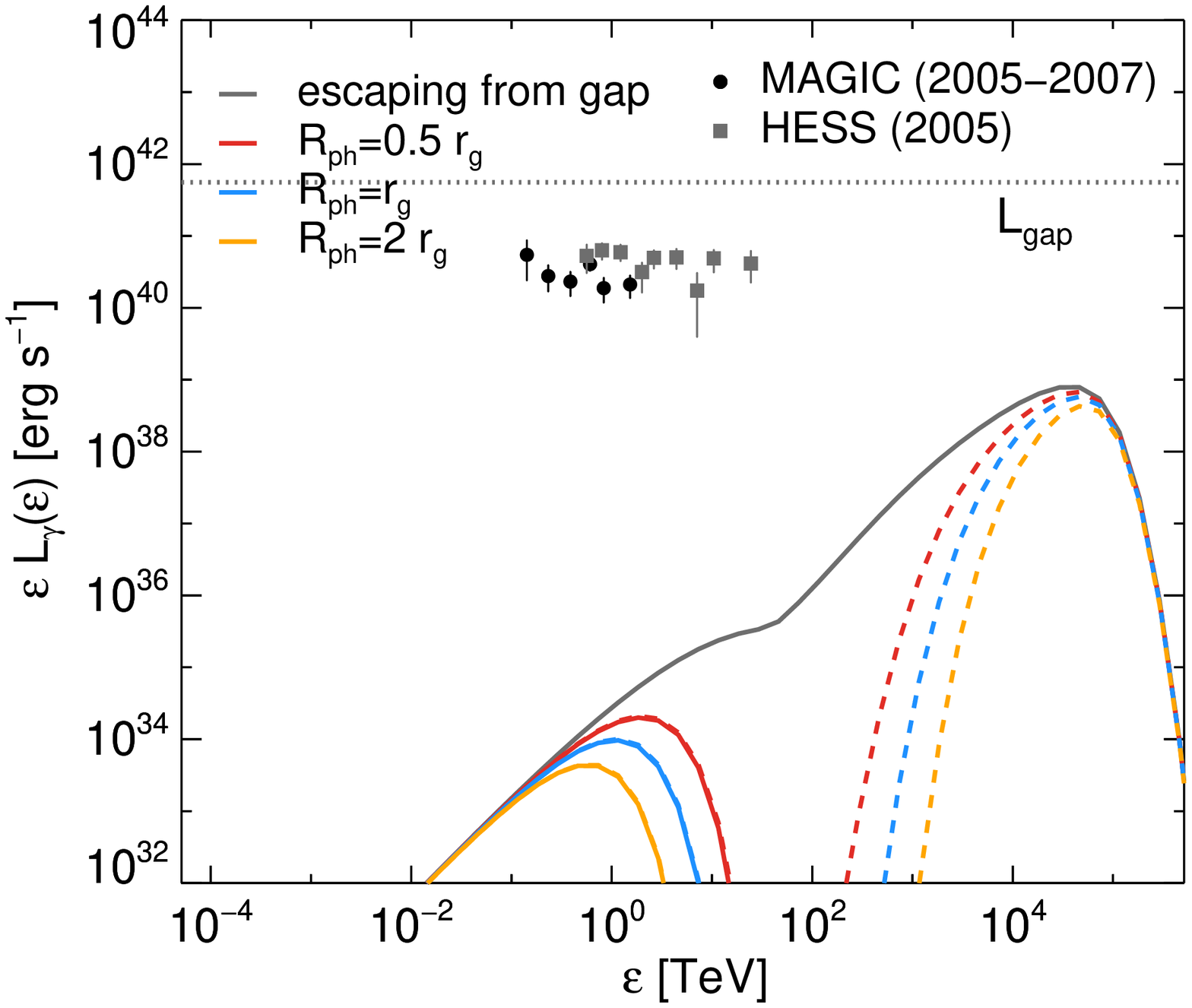}
    \caption{Photon energy spectra escaping from the gap (solid grey lines) for two indicative cases shown in the right panel of \fign{times} with parameters $E_0=5\times10^{-3}$~statV cm$^{-1}$, $h=0.02 \, r_g$ (left panel) and $E_0=9\times10^{-1}$~statV cm$^{-1}$, $h=0.2 \, r_g$ (right panel). Dashed colored lines show the attenuated gamma-ray spectra due to photon propagation in the ambient photon field, assuming different radii $R_{ph}$ (see inset legends). Solid colored lines show the observed spectra after EBL attenuation for the M87 distance. Here, the model of \citet{franceschini08} was adopted. In both panels, the estimated gap power (see \eq{Pgap}) is indicated by the horizontal dotted line. For comparison, we overplot  two indicative spectra of M87 taken with MAGIC during a low state between 2005 and 2007 \citep{aleksic12} and with H.E.S.S. during a flare in 2005 \citep{aharonian06}. } 
    \label{fig:m87}
\end{figure*}

\subsection{Application to M87}\label{sec:m87}
As an indicative example, we consider the radio galaxy M87 (NGC  4486) located at a distance $\simeq 16.7$~Mpc \citep{Mei_2007}. Its soft photon spectrum peaks at 300 GHz, with a flux around 1~Jy 
(corresponding luminosity $L\sim 10^{42}\,\rm{erg \, s}^{-1}$), 
and an estimated photon index  $\Gamma\sim2.2$ above 300 GHz \citep[see][and references therein]{broderick_2015}. However, the actual amount of the soft radiation in the direct vicinity of the gap is quite uncertain. Following \citet{chen_18}, we adopt the 300~GHz frequency as the low-energy end of the power-law photon distribution, i.e., $\epsilon_{\min}=1.2\times 10^{-3}$ eV$=2.4\times 10^{-9}m_ec^2$, and use $U_{ph}=L/(4\pi r^2c)\sim 0.1\,\rm{erg}\,\rm{cm}^{-3}$ as an estimate of the soft photon energy density at a distance $r\approx 3 r_g$ from the black hole. The pair cascade growth rate for these parameters is shown by the red curve in the right panel of Fig.~\ref{fig:times}. The screening of the gap occurs in the  low-$E_0$ (Thomson) regime (see crossover of the red and dashed grey lines) and the  peak electric field  is predicted to be $E_0\sim 5\times10^{-3}\,\rm{statV}\,\rm{cm}^{-1}\sim 5\times10^{-5}\,B_0$ for $B_0=100$~G, consistent with the estimation by \citet{chen_18}. In reality, the properties of the radiation field at the gap (e.g., $U_{ph}$ and $\epsilon_\mathrm{min}$) are highly uncertain. For example, if $\epsilon_\mathrm{min} \gtrsim 10^{-2}\,\mathrm{eV}$ or $U_{ph}\lesssim 10^{-3}\,\rm{erg}\,\rm{cm}^{-3}$, then the pair cascade would be regulated by triplet pair production and Klein-Nishina scatterings. Even though the mean free path for ICS scatterings in the Klein-Nishina regime is expected to be large, it is still possible to screen the electric field in the gap due to an extra source of pairs provided by the TPP process, an effect which has not been considered in the past for M87.

We next discuss the predictions of our gap model about the gamma-ray emission. To do so, we compute the escaping photon spectra for two indicative cases where the screening of the gap happens either in the low-$E_0$ regime (case A)  or in the high-$E_0$  regime (case B). More specifically, we use the parameters that correspond to the intersection of the grey dashed line with the red-colored and blue-colored curves in the right panel of \fign{times}. As discussed previously, the intersection of the curves provides an estimate of the gap size and power, which for cases A and B read respectively $h\sim 0.02 \, r_g$, $P_{\rm gap}\simeq 3 \times 10^{38}$~erg s$^{-1}$  and $h\sim 0.2 \, r_g$, $P_{\rm gap}\simeq 6\times10^{41}$~erg s$^{-1}$.

The photon spectra escaping from the gap for the two cases are presented in \fign{m87} (solid grey lines). When the gap is screened in the low-$E_0$ regime (see left panel), the growth of the pair cascade takes place mostly in the Thomson regime, where pairs cool efficiently via ICS. As a result, the escaping gamma-ray luminosity is comparable  to the gap power and the photon spectra typically extend from tens to hundreds of TeV depending on the parameters. For case A, in particular, we find $L_{\gamma} \approx P_{\rm gap}/2$ and a peak photon energy of $\sim 10$~TeV   (see left panel). On the contrary, if the gap is screened in the high-$E_0$ regime, where Klein-Nishina scatterings and TPP are important, the photon spectrum can extend to much higher energies (e.g., $\gtrsim$~PeV), because electron cooling is suppressed  (see right panel). This also means that a large fraction of the gap power that is being ``pumped'' into the pairs by the electric field is not radiated away. In this particular example, we find that $L_{\gamma}\sim 0.01 \, P_{\rm gap}$. Thus, in both cases, the direct gamma-ray emission from the gap cannot explain the TeV observations of M87 (see symbols in \fign{m87}), but for different reasons: either the gap is screened in the Thomson regime and its power is low (left panel) or the gap is screened in the Klein-Nishina-TPP regime and its power is higher, but the radiative efficiency is low.   

The gamma-ray spectrum that escapes from the gap is not necessarily the same as the observed one. Photons still have to propagate through the ambient radiation field, and the degree of the gamma-ray flux attenuation will depend on the typical size of the radiation $R_{ph}$, as illustrated in \fign{m87} (see dashed colored lines in both panels). Gamma-ray photons with energy $\sim 3/\epsilon_{\min}$ will have the shortest mean free path, and a large dip in the spectrum is expected at this energy, as shown in the right panel of the figure. Photons with either much lower or higher energy will suffer less attenuation. 

The absorbed gamma-rays will initiate a post-gap pair cascade, and the absorbed luminosity will re-emerge at lower energies with a spectrum that will depend on the interplay of various cooling processes (e.g., curvature, synchrotron, ICS in the Thomson or Klein-Nishina regimes). Although the attenuated luminosity, for both cases considered here, will be a small fraction of the observed TeV luminosity of M87, the outflowing pairs from the gap will still carry most of the gap power in case B. Thus, if these ultra-relativistic pairs are able to lose their energy in the post-gap region via channels other than ICS and TPP (e.g., curvature radiation), a radiative signal with $L_{\gamma}\sim P_{\rm gap}$ is expected. A detailed study of the post-gap cascade emission will be presented elsewhere.

\begin{figure}
    \centering
    \includegraphics[width=0.47\textwidth]{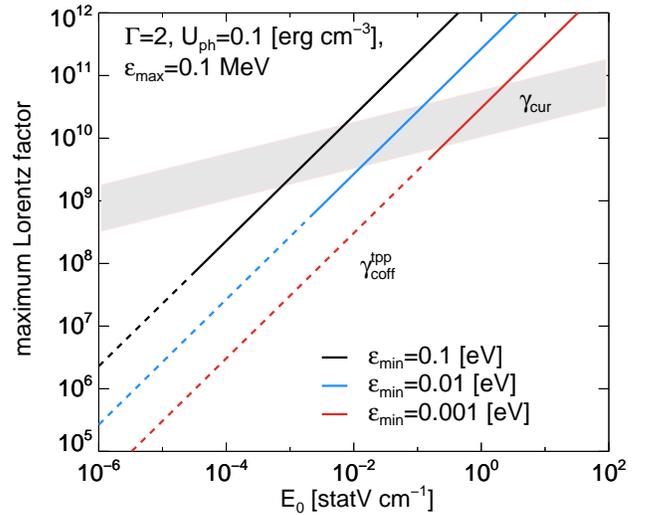}
    \caption{Limiting Lorentz factor due to energy losses by curvature radiation (see \eq{gcur}) as a function of the electric field, for curvature radii ranging from 1 to 30 $r_g$ of a black hole with mass $M_{\rm BH}=6.5\times10^9\, M_{\odot}$ (grey-colored band). Solid colored lines show an estimate for the cutoff Lorentz factor of the pair distribution (see \eq{gtpp}) in the regime where TPP dominates the production of pairs, for different values of the minimum energy of the ambient power-law photon field (see inset legend; same as in the right panel of \fign{times}).  The expression given by \eq{gtpp} is not valid in the low-$E_0$ regime (dashed lines). A colored version of this plot is available online.} 
    \label{fig:gcur-gtpp}
\end{figure}
\section{Discussion}\label{sec:discussion}
We have presented numerical calculations of  pair-Compton cascades in the Thomson and Klein-Nishina regimes and investigated  the effects of triplet pair production. By adopting a simplified model for the acceleration region and using the method of transport equations for the particle evolution, we computed the growth rate of the pair cascade as a function of the accelerating electric field in the presence of black-body and power-law ambient photon fields. Informed by our numerical results, we derived simple analytical scalings of the growth rate (see Eqs.~\ref{eq:s-low} and \ref{eq:smax}) and the corresponding electric field (Eq.~\ref{eq:Ekn}) on the properties of the background photon field. For certain parameters, which can be realized in the vicinity of AGN black holes, the pair cascade may well be regulated by triplet pair production and Compton scatterings in the deep Klein-Nishina regime.

Even if the physical conditions in the acceleration region (i.e., background photon field and electric field strength) are such as to ensure the development of the pair cascade in the deep Klein-Nishina regime, this may not be realized if pairs can lose energy via a process other than TPP and ICS.  Because of the ultra-relativistic energies involved in the high-$E_0$ regime (see right panel in \fign{s-BB}), the most relevant energy loss process becomes curvature radiation. The limiting Lorentz factor in this case is:
    \eqb 
    \label{eq:gcur}
    \gamma_{\rm cur} \simeq 10^{10} \,  E_{0,2}^{1/4} M_{9}^{1/4} \left(\frac{\rho_c}{r_g}\right)^{1/2}
    \eqe 
where $\rho_c$ is the typical radius of curvature for a particle trajectory, which is assumed to be independent of the particle energy. In the high-$E_0$ (TPP-dominated) regime (i.e., for $E_0 \gtrsim E_0^{(\rm KN)} \simeq 5\times10^{-4} \, U_{ph,0} \,\langle \epsilon\rangle_0^{-2}$~statV cm$^{-1}$), the  cutoff Lorentz factor of pairs can be approximated by\footnote{The logarithmic dependence of the TPP cross section is neglected when deriving this expression.}:
\eqb 
\gamma_{\rm coff}^{(tpp)} & \approx &  2\times10^{13} \, E_{0,2} \, \frac{\epsilon_{\min,-1}}{U_{ph,0}} \ln\left(\frac{\epsilon_{\max,5}}{\epsilon_{\min,-1}}\right) \nonumber \\ 
& \approx & 1.4 \times 10^{13} E_{0,2} \langle \epsilon\rangle_0 U_{ph,0}^{-1}
\label{eq:gtpp}
\eqe 
where  the notation $\epsilon_{j,x}=\epsilon_{j,x}/(10^x \, {\rm eV})$ was introduced, a power-law photon field with $\Gamma=2$ and $\epsilon_{\max}\gg \epsilon_{\min}$ was assumed, and \eq{mean} was used to derive the second expression. 
Curvature losses can be neglected as long as $\gamma_{\rm coff}^{(tpp)} \lesssim \gamma_{\rm cur}$ or equivalently:
\eqb 
\label{eq:gtpp-gcur}
\langle \epsilon \rangle_0 \, U_{ph,0}^{-1} \lesssim 7\times10^{-4} E_{0,2}^{-3/4} M_9^{1/4}  \left(\frac{\rho_c}{r_g}\right)^{1/2}.
\eqe 
\fign{gcur-gtpp} shows the two characteristic Lorentz factors as a function of $E_0$ for different values of the minimum energy $\epsilon_{\min}$ of a power-law photon field with index 2. If, for a given $E_0$, $\gamma_{\rm coff}^{(tpp)}>\gamma_{\rm cur}$,  then the actual growth rate of the pair cascade will be lower than the one obtained in \sect{results}. For the indicative parameters of the photon field in M87 that we adopted in the previous section (i.e., $U_{ph}=0.1$~erg cm$^{-3}$ and $\epsilon_{\min}=10^{-3}$~eV) we find that $\gamma_{\rm coff}^{(tpp)}$ becomes comparable to $\gamma_{\rm cur}$ only for $E_0\sim 0.1-1\,\mathrm{statV}\,\mathrm{cm}^{-1}$ (see red line in \fign{gcur-gtpp}). However, our simplified model predicts that the electric field will not be able to grow to such values and that the gap will be screened by the pair growth in the Thomson regime, as shown in \fign{times} (see red curve in right panel). Nevertheless, for other photon field parameters the pair cascade may develop in the TPP-dominated regime before the particle energy is limited by curvature losses (see e.g., black and blue lines in \fign{gcur-gtpp}).

In this work, we studied the pair cascade under the assumption of a time-independent electric field. This is a good approximation, if the particle acceleration timescale is much shorter than the timescale for the electric field to evolve. In a dynamic gap, if the electric field grows quadratically with time, as in \eq{tE}, the condition $t_{\rm acc}(\gmax)<t_E$ in the low-$E_0$ regime can be written as:
\eqb
\frac{E_0}{B_0}>\left(\frac{3}{8\pi}\right)^{1/2}\left(\frac{r_B}{r_g}\right)^{1/2}\left(\frac{\ell_{\rm IC}^{\rm tot}}{r_g}\right)^{1/2}\left(\frac{m_e c^2}{\langle \epsilon\rangle}\right)^{1/2}\!\!\!\!,
\label{eq:tacc-tE}
\eqe 
where we used \eq{gmax}, $r_B=m_ec^2/(q_eB_0)\approx17\,B^{-1}_{0,2}$~cm and $\ell_{\rm IC}^{\rm tot}=\langle\epsilon\rangle/\sigma_TU_{\rm ph}$ is the mean free path for IC scattering on the overall soft photon energy distribution. For $U_{\rm ph}$ and $B_0$ values typical for the black-hole environment of AGN, \eq{tacc-tE} becomes $E_0/B_0>10^{-5}M_9^{-1}B_{0,2}^{-1/2}U_{\rm ph,0}^{-1/2}$, and can be easily satisfied. In the high-$E_0$ (TPP-dominated) regime, the condition on the timescales reads $t_{\rm acc}(\gamma_{\rm coff}^{(tpp)})<t_E$ or equivalently:
\eqb 
\frac{E_0}{B_0} \gtrsim 0.01 \, M_9^{-1} U_{ph,0}^{-2} \langle\epsilon\rangle_0^{2},
\eqe 
where we used \eq{gtpp} and considered a power-law soft photon field with $\Gamma=2$. This condition is more restrictive than the one in \eq{tacc-tE}, since the pairs in the TPP-dominated regime reach ultra-relativistic energies and thereby have much longer acceleration timescales. It is therefore possible that the assumption of  constant electric field breaks down in the  high-$E_0$ regime for certain parameters. In the case where the acceleration time is no more negligible, the growth of the electric field and its response to pair creation should be taken into account. Moreover, our simplifying assumption of identical particles, which disregards the deceleration of positrons, is  questionable. In order to simulate the cascade in this case, a proper particle-in-cell (PIC) treatment \citep{levinsoncerutti18,chen_18} with inclusion of both ICS and TPP processes is necessary. This will be carried out in a dedicated future work.

Here, we have studied the development of  pair cascades initiated by $\gamma \gamma$ absorption of $\gamma$-ray photons produced by accelerated electrons in a gap. It is also possible that a non-relativistic proton  enters the charge-starved region and accelerates to high energies \citep{bednarek_99}. In this case, one expects that protons will start producing pairs via the Bethe-Heitler process ($p\gamma \rightarrow p e^{-}e^{+}$) on ambient soft photons with energy $ \epsilon_0$ when they reach a Lorentz factor $\gamma^{(pe)}_{p} \gtrsim 2 m_e c^2/ \epsilon_0$. The pairs will be injected into the gap with relativistic Lorentz factors and a broad energy spectrum \citep{mast_05, Kelner_2008, PM15}. Similar to the pairs produced by $\gamma \gamma$ pair production, the Bethe-Heitler pairs can initiate electromagnetic cascades leading to high multiplicities (i.e., large number of secondary pairs per proton). It is therefore likely that the electric field will be screened before the protons reach the required energy to produce pions and, in turn, neutrinos, as the threshold for photo-meson production ($p\gamma \rightarrow n \pi^{+} (p\pi^0$)) is higher, namely $\gamma^{(p\pi)}_{\rm p} \gtrsim m_\pi c^2/\epsilon_0$, where $m_{\pi}$ is the pion mass. However, if the background photon field has a wide energy spectrum (e.g., power law), then even lower energy protons can interact with sufficiently high-energy photons to produce neutrinos. The computation of the escaping neutrino and $\gamma$-ray fluxes in this scenario, which becomes relevant in light of the recent neutrino detections from the $\gamma$-ray AGN TXS~0506+056 \citep{aartsen_2018a, aartsen_2018b}, requires a dedicated study of the ensued pair cascade.

\acknowledgments
The authors thank the anonymous referee for useful suggestions that helped to improve the manuscript. This research was motivated by discussions at the 2018 MIAPP workshop ``The High Energy Universe: Gamma Ray, Neutrino, and Cosmic Ray Astronomy''. It is supported by the Munich Institute for Astro- and Particle Physics (MIAPP) of the DFG cluster of excellence ``Origin and Structure of the Universe''. M.P. and Y.Y. acknowledge support from the Lyman Spitzer, Jr.~Postdoctoral Fellowship. MP acknowledges support also from the Fermi Guest Investigation grant 80NSSC18K1745. A.C. acknowledges support from NASA grant NNX15AM30G.

\bibliographystyle{aasjournal} 
\bibliography{cascade.bib}
\end{document}